\documentstyle[aps,prb,multicol,graphicx]{revtex}
\draft
\begin{document}
\title{A Theory for High-$T_c$ Superconductors 
Considering Inhomogeneous Charge Distributions}
\author{E. V. L. de Mello, E. S. Caixeiro and J. L. Gonz\'alez }
\address{Departamento de F\'{\i}sica,
Universidade Federal Fluminense, av. Litor\^ania s/n, Niter\'oi, R.J.,
24210-340, Brazil}
\date{Received \today }
\maketitle
\begin{abstract}
We propose  a general  theory for  the  
critical and pseudogap temperatures $T_c$ and $T^*$  dependence on
the doping concentration for high-$T_c$ oxides, taking into account
the charge inhomogeneities in the $CuO_2$ planes. 
Several recent experiments have revealed that  
the charge density $\rho$  in a given compound (mostly underdoped)
is intrinsic inhomogeneous with large spatial variations which leads 
to a local charge density $\rho(r)$. These 
differences in the local charge concentration  yield insulator
and metallic regions, either in an intrinsic  granular or 
in a stripe morphology. 
In the metallic region, the inhomogeneous charge
density produces also spatial or local distributions  
which form Cooper pairs at a local superconducting critical 
temperatures  $T_c(r)$ and zero temperature
gap $\Delta_0(r)$.   For a given compound, 
the measured onset of vanishing gap temperature is identified 
as the pseudogap temperature, that is,
$T^*$, which is the maximum of all $T_c(r)$. Below $T^*$, due to
the distribution of $T_c(r)$'s, there
are some superconducting regions surrounded by insulator or metallic medium.
The transition to a coherent superconducting state corresponds to the
percolation threshold among the superconducting regions 
with different $T_c(r)$'s. The charge inhomogeneities have been studied
by recent STM/S experiments which provided a model for our
phenomenological distribution. 
To make definite calculations and compare with the experimental results,
we derive phase diagrams for the BSCO, LSCO
and YBCO families, with a mean field theory
for superconductivity using an extended Hubbard Hamiltonian. 
We show also that this novel approach provides new insights on 
several experimental features of high-$T_c$ oxides. 

\end{abstract}

\begin{multicols}{2}
\section{Introduction}
 It is well known that the properties of high-temperature
superconductors (HTSC) vary in
an unusual way when  a moderate density of holes are introduced
into the $CuO_2$ planes by chemical doping. 
This is one of the reasons why, despite
of a large experimental and theoretical effort, 
the nature of the superconductivity in these materials  
remains to be explained\cite{Kitazawa}. 
Correlations make the  parent undoped compound to be
a Mott insulator and, upon doping,  the underdoped compounds display
unusual metallic properties with increasing $T_c$. Doping beyond the optimal
level yields normal metals
with  Fermi liquid behavior and with decreasing $T_c$. 

The non usual properties of underdoped 
samples have motivated several experiments and two features
have been discovered which distinguish them from the overdoped compounds: 
firstly, the appearance of a pseudogap at a temperature $T^*$, 
that is, a discrete structure
of the energy spectrum above $T_c$, identified by
several different probes\cite{TS}. $T^*$ is found to be present
also in overdoped samples\cite{TS,Renner}, but at temperatures
near $T_c$. Second, there are increasing
evidences that the electrical charges are highly inhomogeneous 
up to (and even further) the optimally 
doped region\cite{Egami96,Billinge00,Buzin00,Fournier,Pan,Davis}.
These charge inhomogeneities are neither due to impurities nor
to crystal defects, but are 
intrinsic to the type of cuprate, producing local
lattice distortions in the $CuO_2$ bond length\cite{Egami}. 
In fact, such intrinsic inhomogeneities  are also consistent
with the presence of charge domains either in a granular
\cite{Fournier,Pan,Davis} or 
in a stripe \cite{Tranquada,Bianconi,EKT} form. 

Therefore, in our view, it is very likely
that  the pseudogap and the intrinsic 
charge inhomogeneities are closely related  
and understanding their interplay is of great 
importance to understand the general 
phenomenology and the phase diagrams of the HTSC families.
Currently there are two main different proposals to explain the existence
of the pseudogap: In the first one, 
the pseudogap is regarded as a normal state 
precursor of the superconducting gap due to local dynamic pairing 
correlations in a state without long range phase coherence, and $T_c$
is much smaller than $T^*$ because of strong phase 
fluctuations\cite{Randeria,EmKi}. In the second proposal, the pseudogap
is a normal state gap, which is necessarily independent of the 
superconducting gap
and which competes with the superconductivity, existing even below
$T_c$ for compounds around the optimum doping,
ending in a quantum critical point\cite{Tallon,Loram,Williams} at
zero temperature.

Recently we have described  a simple new scenario\cite{Mello01}:
when the temperature of a given sample is decreasing and reaches
$T^*$,  it  activates the formation of 
pairs  at some selected low doping metallic regions. 
Initially these superconducting or pair-rich regions
are not connected and there is no phase coherence over the 
whole system.
The coherent superconducting transition occurs when the temperature reaches
a value ($T_c$) at which the different superconducting regions percolate. 
This scenario relies heavily upon the fact that the cuprates
have an intrinsic inhomogeneous charge distribution. Thus
a given HTSC compound with an average hole per $Cu$ ion
density  $\langle \rho \rangle$
and with an  inhomogeneous microscopic charge distribution $\rho(r)$
has a distribution of small clusters or stripes  with a given local $T_c(r)$
($0< \langle \rho \rangle \le 1$ and $0< \rho(r) \le 1$).
Thus the model depends strongly on the intrinsic charge
distribution $\rho(r)$ although their exactly form  is not well known and
since the discovery of the spin-charge stripes\cite{Tranquada},
they  are a matter of intense current research. 
As a consequence, several recent experiments 
have demonstrated that the charge distributions are 
more inhomogeneous for underdoped and more homogeneous 
for overdoped compounds and may resemble either a granular
\cite{Egami96,Billinge00,Buzin00,Fournier,Pan,Davis} or
a stripe structure.

In the spin-charge striped scenario, some regions of the plane are heavily
doped (the stripes) and other regions are underdoped, filling the space
between the charge-rich stripes.
Stripe phases occur due to the
antiferromagnetic interaction among magnetic ions  and Coulomb interaction
between the charges, both of which favor localization. On the
other hand, the zero-point motion of the holes favors
delocalization and tends to create phase-separated states rich
either of spin or charge.
Experimentally, the stripes are more easily
detected in insulating materials, where they are static, but there
are evidences of fluctuating stripe correlations in metallic and
superconducting compounds\cite{Bianconi,Tranquada,EKT}.

Based on the results of the  above mentioned experiments, we 
have introduced a local charge distribution $\rho(r)$
to model the real  charge distributions
inside a HTSC compound. Since doping produces an intrinsic
inhomogeneous charge distribution and metallic and insulator
regions seems to coexist, we
assume that it  contains two parts: one for a hole-rich
and other for a hole-poor partition. This mimics the striped
phases. The
hole-poor regions are, in most cases, antiferromagnetic Mott insulators. 
The hole-rich regions form an inhomogeneous metal with spatially
varying charge density. The existence of the intrinsic charge 
inhomogeneities have several consequences,
and one of them is the non-Fermi liquid behavior of the
underdoped compounds. 
As concerns the superconductivity, they produce spatially dependent
superconducting gaps $\Delta_{sc}(r)$ due to short coherence length
and also spatially dependent  superconducting domains
with critical temperatures $T_c(r)$. 
Recently, very fine scanning tunneling microscopy/spectroscopy
(STM/S) data\cite{Fournier,Pan} has revealed
the spatial variation through the  differential conductance which
provides a strong evidence for such  distribution of  zero temperature
superconducting gap $\Delta_{sc}(r)$. More recently, new STM data
on Bi2212 using scattering resonances at Ni impurities atoms, 
has revealed a large nanoscale spatial dependence of the superconducting
gap\cite{Davis}. 

All together, these experiments show variations on
local density of states (LDOS) of just a few angstroms. This is
consistent with the very short coherence length in HTSC. 
Since $T_c(r)$ is proportional
to  $\Delta_{sc}(r)$, the opening of the largest gap occurs at
the highest of all the $T_c(r)$'s, which is exactly $T^*$.
Above $T^*$ there is no gap and below it, there is the 
development of some superconducting
clusters  inside the material. Depending on the distribution of 
the values of   $T_c(r)$ in the compound,
some regions become superconducting. Decreasing the temperature, the
number of superconducting clusters increases, bringing about the 
superconducting domains to grow. When the temperature reaches a
value which is the $T_c$ of the compound, the
superconducting regions percolate through the sample and therefore
it can hold a dissipationless current. 

On Fig.1 we show a schematic
local charge density distribution $\rho(r)$ which is based on the 
above experimental data, as mentioned, and could give rise to 
a gap distribution
of $\Delta_{sc}(r)$. This is a pictorial sketch of what should be
the real charge distribution for a optimal doped HTSC, based 
on the above information. It reflects the nano-scale variations of the
superconducting energy gap or the nano-scale variations on the
density of charges in the material.

\begin{figure}
\includegraphics[width=8cm]{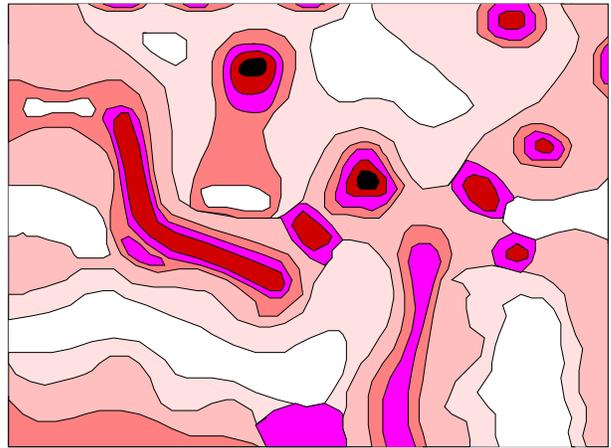}
\caption{Sketch of the possible charge distribution
in a HTSC compound. The white part represents the insulator or
hole-poor regions. The colored region mimics the hole-rich metallic
part of the compound. The lighter colors are for metals with higher
values of the superconducting gap and the darker regions is for lower 
gap regions.}
\end{figure}

This percolating scenario can be understood by analyzing
the scanning SQUID microscopy magnetic data which 
makes a map of the expelled magnetic flux (Meissner effect) domains 
on LSCO films\cite{IYS}. This experiment  shows the regions where the 
Meissner effect continuously develops from near $T^*$ to temperatures
well below the percolating threshold $T_c$. 
Notice that the regions without magnetic flux due to the 
Meissner effect can exist only where
the Cooper pairs are in  phase coherence in a large spatial region. 
Furthermore there is no apparent difference, other than the size of
the superconducting region, as the temperature changes across $T_c$.
Secondly, it shows the existence
of the superconducting regions above $T_c$ which differ only in
size from those below $T_c$. Furthermore,
the d-wave nature of both pseudogap and order parameters, the 
tunneling conductance and ARPES measurements\cite{TS} with their curves
evolving smoothly across $T_c$, are compelling evidences in 
favor of a similar nature of both gap and pseudogap. 

Several different ideas of a superconducting phase in cuprates attainable
by some form of percolation between pairs, cluster or stripes
are not new\cite{HKS,Muller,Mesot}. More recently,
a percolating  approach was suggested and  supported by 
the microscopic model based on a spatial
dependent critical temperature $T_c(r)$ due to  
intrinsic inhomogeneities like  
pair breakers or spatial dependence of the superconducting
coupling constant proposed by Ovchinnikov et al\cite{OWK}. They
have derived the density of states due to the spatial 
distribution of coupling constants.  A different method 
based on the bond percolation
between  pre-formed mesoscopic Jahn-Teller pairs was proposed
recently\cite{Mihailovic}
as a mechanism to attain  a superconducting phase coherence . 
On the other hand, the spatial variations
on the LDOS due to charge disorder were taken into account in 
calculations using a generalized $t-J$ model\cite{Wang}. In fact,
new trends an ideas have been used to describe the effects of 
the intrinsic disorder and
the possibility of a percolating phase not only for
HTSC but also for manganites\cite{Burgy}.

The ideas described in the above paragraph are on the
same lines with the scenario that we want to discuss to
explain why the occurrence of pairing and the phase coherent
superconducting state in the  high-$T_c$ superconductors appears 
to occur independently:
we show that the percolation theory, i.e., that the
superconducting state is reached through the percolation
of regions rich in pre-formed pairs which provides 
good quantitative agreement with the  measured
$T^*$ and $T_c$  phase diagrams.
Notice that there is a general agreement with respect
to the experimental $T_c (\langle \rho \rangle)$ 
curves, since most of them are obtained
through the same method, namely, resistivity measurements. 
However the measured values of $T^*$ found in the 
literature are obtained through different
methods and seems to vary considerably, depending on the specific
experimental probe used\cite{TS}. Such difference may be due to the anisotropic
d-wave nature of the gap amplitude and the fact that a given
experimental technique is sensitive to excitations at a particular
wavevector magnitude. Therefore, it is natural that angle-resolved
photoemission (ARPES), tunneling spectroscopy, transport properties
such as dc resistivity and optical conductivity, NMR,  Knight shift
relaxation rate, electronic Raman, magnetic neutron scattering,
specific heat\cite{TS,Tallon} and recent vortex-like Nernst signal
measurements\cite{Xu} yield different values for the 
onset of vanishing gap temperature $T^*$. Since the origin of
the charges inhomogeneities are not known, $T^*$ could be 
also dependent on the way a given sample is made.

We must also emphasize that the percolating approach is not only suitable to 
yield good quantitative agreement with the
HTSC phase diagrams but, and perhaps more important, it
provides interesting new physical insight on a number of phenomena detected
in these materials: the variation of the measured pseudogap magnitude 
with the  temperature\cite{TS}, the decreasing of the zero
temperature superconducting gap $\Delta_0$ while $T_c$ increases
for underdoped compounds\cite{Renner,Harris}, the downturn of the linear
dependence of the resistivity with the temperature for underdoped
samples and the increase of Hall carriers with
the temperature, mostly measured for the 
optimally doped and underdoped compounds\cite{Ong}, the downturn of the linear
specific heat coefficient,.... These properties will be discussed
in more detail in section IV.

This paper is divided as follows: in section II, we introduce the 
charge distributions appropriate  to mimic
the real charge distribution inside the material  
and reproduce the $T^*$ and $T_c$ phase
diagram. In section III, we derive the phase diagram
for Bi2212, LSCO and YBCO and make comparison with the 
experimental data. We use a mean field BCS like method with
an extended Hubbard Hamiltonian to derive the onset of vanishing
gap at $T^*$, but it should be emphasized that the percolating approach  
introduced here is independent of any method of calculating the
superconducting pair formation. In section IV, we comment on the applications
and implications of the percolation theory to
several physical properties of the HTSC and we finish with
the conclusions in section V.
\section{The Charge Distribution}
The consequence  of the microscopic charge inhomogeneities distribution
in the $CuO_2$ planes, either in a striped or in a granular configuration,
is the existence of domain walls between the
two phases which are spontaneously created in the 
planes\cite{Tranquada,Bianconi,EKT}:
regions which are heavily doped or hole-rich may form
the stripes and others regions which are hole-poor 
are created between these charge-rich stripes. The exactly
form of these charge distributions is not well known but we
can get some insights from a number of recent experiments.  
The neutron powder diffraction\cite{Billinge00,Buzin00,Egami} suggests
that the charge inhomogeneities modifies the $Cu-O$ bond length, leading
to a distribution of bond lengths for optimal and underdoped compounds.
Scanning tunneling 
microscopy/spectroscopy (STM/S)\cite{Fournier,Pan} on optimally doped
$Bi_2Sr_2CaC_2O_{8+x}$ measures nanoscale spatial variations  
in the local density
of states  and the superconducting gap at a very short length 
scale of $\approx 14 \AA$. These results suggest that instead
of a single value, the zero
temperature superconducting gap assumes different values at different
spatial locations in the crystal and their statistics
yield a Gaussian distribution\cite{Pan}. New high resolution STM
measurements\cite{Davis} have revealed an interesting map of 
the superconducting gap spatial variation for underdoped Bi2212. 
Their data is compatible, as they pointed out, with a granular
superconducting grains separated by non-superconducting regions.
Based on their measured gap histogram (see Fig.4b of Lang
et al\cite{Davis}) and on similar data of Pan et al\cite{Pan},
we can draw some insights on the real charge distribution
inside a HTSC.

Thus, in order to model the above experimental observations and 
to be capable of performing calculations which may  
reproduce the measured phase diagrams, namely $T^*$ and $T_c$ for
a given family of compounds, 
we used a combination of a Poisson and a
Gaussian distribution for the charge distribution $\rho(r)$.  
In fact, each type of distribution
has a convenient property which we used below: 
The width of a  Gaussian distribution is easy to be controlled
and  expresses the degree of disorder but since it is symmetric,
it would imply in a rather small width to the low density compounds. 
On the other hand, a Poisson distribution
starts sharply and has a long tail which is convenient to deal with these
these low densities compounds with their experimental measured
large degree of inhomogeneities. Thus,
for a given compound  with an average charge density $\langle \rho \rangle $, 
the hole distribution $P(\rho; \langle \rho \rangle)$ or simply
$P(\rho)$ is a  histogram  of the  probability of
the local hole density $\rho$ inside the sample,
separated in two branches or domains.
The low density branch represents the  hole-poor or non-conducting regions and
the high density one represents the  hole-rich or metallic regions.
As concerns the  superconductivity, only the properties of the hole
rich branch are important since  the current  flows only through the
metallic region. 

Such normalized charge probability distribution may be given by:

\begin{eqnarray}
 P(\rho) &=&  (\rho_c-\rho)\exp[-(\rho-\rho_c)^2/2(\sigma_-)^2]/ \nonumber
       [(\sigma_-)^2(2-\exp 
\\ &&(-(\rho_c)^2/2(\sigma_-)^2))] \quad\text{ for }\quad 0<\rho<\rho_c
\label{equationa}
\\
P(\rho) &=& 0  \quad\text{ for }\quad \rho_c<\rho<\rho_m
\label{equationb}
\\
P(\rho) &=&  (\rho-\rho_m)\exp[-(\rho-\rho_m)^2/2(\sigma_+)^2]/ 
       [(\sigma_+)^2(2-\exp        \nonumber
  \\ &&(-(\rho_c)^2/2(\sigma_-)^2))] \quad\text{ for }\quad \rho_m<\rho
\label{equationc}
\end{eqnarray}

The values of $\sigma_-$ ($\sigma_+$)controls the width of the low (high) 
density branch. Here, $\rho_c$ is the end
local density of the hole-poor branch. $\rho_m$ is 
the starting local density of the hole-rich or 
metallic branch. Both $\rho_c$ and $\rho_m$ are shown in Fig.2 for
a particular case. Each
compound will have a its value of $\rho_m$.
Thus, a  compound having  regions with different hole concentrations
which are distributed according to the above 
probability density function $P(\rho)$,
has an average value which depends on the parameters of the distribution.
We can show that the average value of the charge density is

\begin{eqnarray}
 \langle \rho \rangle &=& [\sigma_+sqrt(\pi/2)+\rho_m+\rho_c- 
 \sigma_-sqrt(\pi/2)\times 
 \nonumber    \\&&   \text {erfunction} (\rho_c/sqrt(\pi/2))]/  \nonumber
          \\&&  [(\sigma_+)^2(2-\exp(-(\rho_c)^2/2(\sigma_-)^2))].
\label{mean}
\end{eqnarray}
here erfunction is the error function. 
For most compounds, due to
the small values of $\sigma_+$,  $\langle \rho \rangle \approx \rho_m$, see
Table I below. Indeed, for a compound with average density $\langle \rho
\rangle $,  
the values of $\sigma_{\pm}$are chosen in order that
percolation in the hole-rich branch occurs  at
a given density called $\rho_p$. 

As the doping level $\langle \rho \rangle $ increases, the
compounds become more homogeneous and therefore,
we take larger values of  $\sigma_-$ and smaller values of the
metallic branch width $\sigma_+$. We take $\rho_c\approx 0.05$ as the 
end of the 
low branch because it corresponds to the onset of superconductivity to
mostly cuprates. However, $\rho_c\approx 0.1$ for the Bi2212 family. 
During the preparation of this work, we discovered that the 
charge probability distribution for the metallic branch, 
introduced as an ansatz in the beginning of our  
calculations, but based on the results of Pan et al\cite{Pan}, 
was confirmed  by the very fine STM measurements of
Lang et al\cite{Davis}. They measured the gap probability  
distribution which has a Gaussian form but with a long tail which is 
more characteristic of a Poisson type distribution. Since there is
not any experimental measurements for the hole-poor branch, $\sigma_-$
is just a free parameter.

\begin{table}[!ht]
\begin{center}
\begin{tabular}{|c|r|r|r|r|r|} 
 & $\rho_m $ & $\sigma_-$ & $\sigma_+$ & $\langle \rho \rangle$ & $\rho_p$ \\
\hline  & 0.080*      & 0.033      & 0.050      &  0.09    & 0.245 \\  
 & 0.100       & 0.034      & 0.050      &  0.10    & 0.240 \\  
 & 0.120       & 0.037      & 0.050      &  0.12    & 0.238 \\  
 & 0.160*      & 0.057      & 0.040      &  0.16    & 0.230 \\  
 & 0.200       & 0.078      & 0.026      &  0.20    & 0.239 \\  
 & 0.220       & 0.090      & 0.024      &  0.22    & 0.245 \\  
 & 0.260*      & 0.04(G)    & 0.04(G)    &  0.26    & 0.268 \\  
\end{tabular}
\caption{Some selected parameters for different doping level 
compounds. Notice that
they are characterized by the average hole density $\langle \rho \rangle $ 
given in
the fourth column. Notice also that the values of the metallic branch
width $\sigma_+$ decreases with $\langle \rho \rangle $, showing that the degree
of disorder decrease with the doping. The $*$ marks the distributions
plotted in Fig.2. The value of $\sigma_+=0.04$ for $\langle \rho 
\rangle =0.16$ is
taken from the experimental data of Pan et al\cite{Pan}.}
\label{table}
\end{center}
\end{table}

We can get a reliable estimation of the $ \sigma_+$ values from 
the experimental STM/S Gaussian histogram  distribution
for the local gap\cite{Pan} in an  optimally doped 
$Bi_2Sr_2CaCu_2O_{8+x}$ ($\rho_m \approx 0.32$). From Fig.2d of Ref.7,
one sees that the half-width for the metallic branch
is about half of the mean doping value, that is
$2\sigma_+=0.16$,
since the local gap is proportional to the local hole density. 
As mentioned, the metallic branch distribution has exactly
the same form of the STM histogram data of Lang et al\cite{Davis}
which express the inhomogeneous superconducting regions in Bi2212. For
the LSCO system, assuming that it has the same  STM/S Gaussian histogram 
distribution, the optimally doped 
compound ($\langle \rho \rangle \approx 0.16$) must have $2\sigma_+=0.08$.
Since the $ \sigma_+$ values are related with the
degree of inhomogeneities, we estimate the $ \sigma_+$  values
for the other compounds using the fact that it decreases with
the hole density. Below we show some selected distributions
for an underdoped, optimally doped and overdoped compounds.

\begin{figure}
\includegraphics[width=8cm]{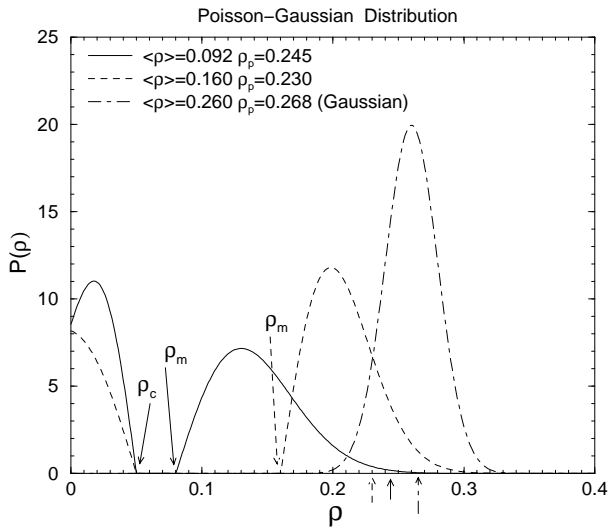}
\caption{Model charge distribution  for
the inhomogeneities or stripe two phase regions. The low density
insulating (antiferromagnetic) branch is near $\rho=0$. The high
density hole-rich region starts at the compound average density 
$\rho_m $ which are indicated by long arrows for some selected
compounds. $\rho_p$, indicated by the short arrows,
is the density where percolation can occur.}
\end{figure}

The density distribution is normalized to unity. 
For compounds like LSCO with average densities
of $\rho \gg 0.25$, that is, in the far overdoped region, we use
a single Gaussian distribution which reflects the more homogeneous
character of these compounds. For Bi2212, since the optimally 
doped compound is about twice that of the other families, the
single Gaussian distribution is appropriate for compounds
with $\rho$ $\gg$ $0.5$.
In the next section we show how to estimate
$T_c$ as function of the local density $\rho(r)$. As already
discussed, HTSC compounds are intrinsic
inhomogeneous, and each region in the material may have a different 
superconducting critical temperature  $T_c(r)$.
The maximum of these temperatures $T_c(r)$ is the pseudogap
temperature since it is connected with the opening of the
largest gap, that is,
the $T^*$ of a given sample with $\langle \rho \rangle$, 
i.e.  $T^*(\langle \rho \rangle)$.
Eventually for temperatures below $T^*(\langle \rho \rangle)$ the 
regions with local density
between $\rho_m$ and $\rho_p$ becomes superconducting and 
the supercurrent may percolate through the sample.
Therefore, the local $T_c(\rho_p)$ is the maximum
temperature at which  a dissipationless current can flow
through the system and which is identified as
$T_c(\langle \rho \rangle)$, the experimental superconducting critical
temperature. 

According to the percolation theory, the site percolation 
threshold occurs in a square lattice
when 59\% of the sites  are filled\cite{Stauffer}.
Thus, we find the density where the hole-rich branch percolates
integrating $\int P(\rho) d\rho$ from  $\rho_m$ till the integral
reaches the value of 0.59, where we define $\rho_p$. Below  $T_c(\langle
\rho \rangle)$ the system percolates and, consequently,
it is able to hold a dissipationless supercurrent. We should
point out that 0.59 is an  appropriate value for a single layer cuprate.
The site percolation threshold is 0.16 for a simple cubic
lattice\cite{Pike}. Therefore, for 
a two layers system like Bi2212, the site percolation
threshold may be less than 0.59, but since the true value is
not known, we will use 0.59 as a first approximation. In the Table I, we
show some of the parameters used for certain sample and we plot the
distribution, as discussed above, in Fig.2.
Notice that the Table I and Fig.2 are for a HTSC system  with optimum
doping $ \langle \rho \rangle \approx 0.16$ like LSCO or YBCO. For Bi2212, the
optimum hole doping is about twice this value\cite{Konsin} and in this
case new parameters must be used. Some charge distributions used
for the calculations with the Bi2212 
family, similar to the ones shown in Fig.2 above, were also 
derived\cite{Mello01}.

\section{The Phase Diagram }
The experimental based phenomenological charge distribution
probability introduced in the last section enable us to
estimate the appearance of pre-formed pair at different
regions in the crystal provided that we know
the onset of superconductivity $T^*(\rho(r))$
There are several different approaches  which can be used to obtain
$T^*(\rho(r))$. We can simply use the {\it experimental} measured
pseudogap temperature $T^*=T^*(\langle\rho\rangle)$ for several
different compounds
through the identification $T^*(\rho(r))$=$T^*(\langle\rho\rangle)$
which is probably the best estimation.
Another possibility is perform a theoretical calculation as we
have done in the past\cite{Mello96,Angilella1} for the $T_c$ of a
given compound with density
$\langle\rho\rangle$ and take $T^*(\rho(r))=T_c(\langle\rho\rangle)$.

Strictly speaking, due to  the non-uniform charge
distribution, we do not have translational symmetry and
we should use a method which takes the disorder into
account\cite{OWK,Wang,Burgy,Atkinson,Ghosal}. However such
theories requires the
introduction of a phenomenological random potential to
simulate the disorder.
Since our purpose here is to demonstrate that phase coherence is attained
by percolation and it is {\it independent} of the pairing
mechanism, we will take the
simplest theoretical approach\cite{Mello96}: we use a BCS-type mean-field
approximation in $k$-space on a two-dimensional square lattice
with uniform carrier density $\rho$ to estimate the onset temperature
of the superconducting d-wave gap, i.e. $T^*(\rho)$. Then we
take $T^*(\rho(r))=T^*(\rho)$ in
order to make a map of the superconducting region. We have to
bear in mind that this is an
approximation but which is worthwhile since it capable to 
reproduce the experimental $T^*$ values
and is reasonable if the size of the grain
or stripes domains are
larger than the typical pair coherent length. In fact, the
latest STM/S results yields grain boundaries of the order of
100-10 nanometers while the typical coherent lengths are of the order
of angstroms.
Thus, this approach, with appropriate choice of experimental or calculated 
parameters,  was used before 
to derive the $T_c (\rho)$ curves\cite{Angilella1,Mello96,Mello99}
for some HTSC families,
since $T_c$ was, as in normal superconductors, taken as the onset
of vanishing gap and $\rho$ was taken as  $ \langle \rho \rangle$.
A two dimension extended Hubbard
Hamiltonian in a square lattice has been used to model the
quasi-bidimensionality of the carriers motion through 
the $CuO_2$ planes\cite{Angilella1,Mello96,Mello99,Schneider} 
and is given by

\begin{eqnarray}
H&=&-\sum_{\ll ij\gg \sigma}t_{ij}c_{i\sigma}^\dag
c_{j\sigma}+U\sum_{i}n_{i\uparrow}n_{i\downarrow}
\nonumber \\
&& +\sum_{<ij>\sigma
\sigma^{\prime}}V_{ij}c_{i\sigma}^\dag c_{j\sigma^{\prime}}^\dag
c_{j\sigma^{\prime}}c_{i\sigma}, \label{b}
\end{eqnarray}
where $t_{ij}$ is the nearest-neighbor and next-nearest-neighbor hopping
integral between sites $i$ and $j$; $U$ is the Coulomb on-site correlated
repulsion and $V_{ij}$ is the attractive interaction between nearest-neighbor
sites $i$ and $j$.  $a$ is the lattice parameter.

Using the well known BCS-type mean-field
approximation to develop Eq.(\ref{b}) in the momentum space,
one obtains the self-consistent gap equation, at finite
temperatures~\cite{de Gennes}

\begin{equation}
\Delta_{\bf k}=-\sum_{\bf k^{\prime}}V_{\bf
kk^{\prime}}\frac{\Delta_{\bf k^{\prime}}}{2E_{\bf
k^{\prime}}}\tanh\frac{E_{\bf k^{\prime}}}{2k_BT},\label{cc}
\end{equation}
with
\begin{equation}
E_{\bf k}=\sqrt{\varepsilon_{\bf k}^2+\Delta_{\bf k}^2}, \label{rr}
\end{equation}
which contains the dispersion relation $\varepsilon_{\bf k}$, and
the  interaction potential $V_{\bf kk^{\prime}}$ which comes  from the
transformation to the momentum space of
Eq.(\ref{b})~\cite{Angilella1,Schneider}. In the calculations we
have used a dispersion relation derived from the ARPES data\cite{Schabel}
with five neighbors hopping integrals. The hopping integrals 
could also be estimated from band structure calculations~\cite{Hybertsen}.
The interaction potential may be given by\cite{Angilella1,Schneider}.

\begin{equation}
V_{\bf kk^{\prime}}=U+2V\cos(k_xa)
\cos(k_x^{\prime}a)+2V\cos(k_ya) \cos(k_y^{\prime}a). \label{d}
\end{equation}

The substitution of Eq.(\ref{d}) into Eq.(\ref{cc}) leads to appearance of a
gap with two distinct symmetries~\cite{Angilella1}:

\begin{equation}
\Delta_{\bf k}(T)=\Delta(T)[\cos (k_xa)\pm \cos (k_ya)],
\label{GG} 
\end{equation}
where the plus sign is for extended-$s$ wave and the minus sign, for
$d$ wave symmetry. In accordance with Ref~\cite{Angilella1} one observes that
the $d$ wave part of the gap do not depend on the coupling constant $U$,
depending only on $V$. Here we deal only with the $d$-wave 
which is the more accepted pseudogap symmetry\cite{TS}.

Using the same BCS-type mean-field approximation,
one obtains the hole-content equation~\cite{Leggett}

\begin{equation}
\rho(\mu,T)=\frac{1}{2}\sum_{\bf k}\left(
{1-\frac{\varepsilon_{\bf k}}{E_{\bf k}} \tanh\frac{E_{\bf k}}
{2k_BT}}\right), 
\label{f}
\end{equation}
where $0\leq \rho \leq 1$.

Eq.(\ref{f}) together with the gap equation (\ref{cc}) 
must be solved together self-consistently. They may be used to derive the
onset of vanishing gap temperature as function of 
a given value of the density of carriers $\rho$. This procedure was used in
the past to derive, with appropriate set of parameters,
the $T_c (\rho)$ phase diagram for different
HTSC systems for a single type\cite{Angilella1,Mello96,Schneider,Edson} 
and for mixture of different order parameter symmetries\cite{Mello99b} . 

In the
present work, we assume a different view, $\rho$ here is not the
compound average hole density but the {\it local} 
hole density $\rho(r)$. 
For a given value of $\rho(r)$, we use Eq.(\ref{f}) and Eq.(\ref{cc})
to calculate the {\it local} onset temperature of vanishing
gap. It is more appropriate to call this temperature as $T^*(\rho)$;
therefore, $T^*(\rho)\equiv T_c(\rho)$ and hereafter we will
deal only with $T^*(\rho)$. On the other hand, the pseudogap
temperature of the compound is $T^*( \langle\rho\rangle)=$max$\{T_c(\rho)\}$,
or simply $T^*$. 

Below, in Fig.3,  we present the results for the 
$La_{2-x}Sr_xCuO_4$ family. We have used a dispersion relation
derived from the Schabel et al\cite{Schabel}. In their notation,
the hopping parameters are: 
$t\equiv t_1$=0.35eV, $t_2$/$t_1$=0.55,
$t_3$/$t_1$=0.29, $t_4$/$t_1$=0.19, $t_5$/$t_1$=0.06. The magnitude of the
attractive potential was set $V/t$=-0.40 in order to give a 
reasonable agreement with many measured values of $T^*$.
As we have already mentioned, different
experiments yield completely different results for  $T^*$.
The phase diagram depend on the 
hopping parameters and on the attractive
potential, which is the free parameter on this type of calculation.
Thus, varying $V/t$ makes the values of the calculated 
$T^* (\rho)$ change and the
position of the optimal density depends on the hopping
parameters. The  hopping values used below
are within the variation estimated  by different
methods (band structure, ARPES measurement, etc.)~\cite{Raimondi,Andersen}. 

The theoretical curves in Fig.3 are derived in the follow way:
For each local density $\rho > \rho_m$, that is, inside the
metallic branch, $T^*(\rho)$, calculated with the mean field
equations, is a decreasing function. Thus the
maximum value of $T^*(\rho)$ is equal $T^*(\rho_m)$.
This is the onset temperature of the superconducting gap (in the metallic
branch) and therefore $T^*(\rho_m)=T^*(\langle \rho \rangle)$. 
To show this, we draw an
arrow in Fig.3 showing how $T^*(\langle\rho\rangle=0.16)$ is obtained. 
The  $T^* (\langle\rho\rangle)$ curve calculated with the above
parameter is in good qualitative agreement with the displayed experimental
data\cite{Xu,Oda}. 
The value of superconducting critical temperature  $T_c(\langle \rho \rangle)$ 
is estimated in the following manner: we
calculate the  maximum temperature at which the superconducting 
region  percolates
in the metallic branch. This percolation occurs when all the clusters with
local density between $\rho_m$ and $\rho_p$ are superconducting. 
Fig.3a shows how this happens for $\langle \rho \rangle=0.16$ which
yields $\rho_p=0.23$. This value of $\rho_p$ can be seen in the panel
following the arrow which shows that  $T^*(0.23)$ is equal the
superconducting critical temperature of the compound, 
$T_c(\langle 0.16\rangle)$.
In general  $T_c(\langle \rho \rangle)\equiv T^*(\rho_p)$ and
the system will be superconducting when submitted to 
any temperature below $T^*(\rho_p)$ (see Fig.3b).
Below we show the results for the LSCO family.
A similar curve was also
studied which the $T^*$ is in reasonable agreement with the 
$T_{c}^{MF}$ extracted from the high-resolution dilatometry
data\cite{Meingast} for $YBa_2Cu_3O_x$.

\begin{figure}
\includegraphics[width=8cm]{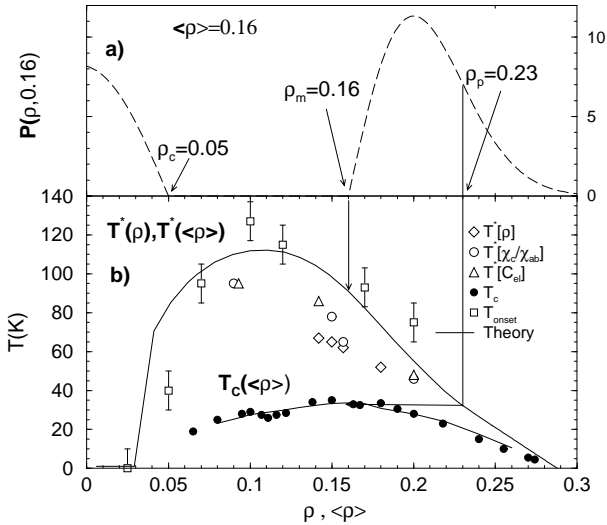}
\label{ladiagr}
\caption{Phase diagram for the LSCO family. To explain how 
$T^*(\langle \rho \rangle)$ and $T_c(\langle \rho \rangle)$  are obtained,
we plot in (a) the probability distribution for the optimal compound
with $\langle \rho \rangle=0.16$, P$(\rho,0.16)$. The arrows shows  $T^*(0.16)$
and the  percolation threshold at  $\rho_p=0.23$ with 
$T_c(\langle \rho \rangle=0.16)= T_c(0.23)$. 
The  experimental data are taken from Ref.~\cite{Oda}
and $T_{onset}$ is taken from the flux flow experiment of Ref.~\cite{Xu}
(open squares). Notice that values of 
$T^*(\langle \rho \rangle)$ are the same of  $T_c(\rho)$ but 
$T^*$ refers to the compound density $\langle \rho \rangle$
and  $T_c$ refers to the local density $\rho$.}
\end{figure}

We have also used this procedure to calculate the phase diagram
for the Bi2212 family which is in agreement with the 
experimental data as one can see in Fig.4. The details of 
the calculations are similar to the ones described above and 
can be found in ref.19.

\begin{figure}
\includegraphics[width=8cm]{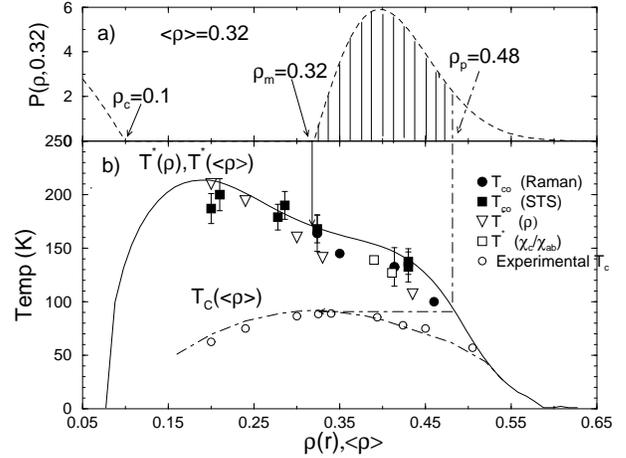}
\caption{a) The probability distribution for the optimal compound
with $\langle\rho\rangle=0.32$. b)The calculated local onset of vanishing
gap $T^*(\rho(r))$ and the compound pseudogap temperature
$T^*(\langle\rho\rangle)$. The thick arrow shows
$T^*(\langle\rho\rangle)$  and the
dot-dash arrow shows how $T_c(\langle\rho\rangle)$ is determined
from $T^*(\rho_p)$.
The experimental points and the symbols are taking from Ref.48.
Notice that values of
$T^*(\langle \rho \rangle)$
refer to the compound average density $\langle \rho \rangle$
and $T^*(\rho)$ refers to the local onset of vanishing gap
at a cluster of density $\rho$.}
\end{figure} 

It is worthwhile to mention that the percolating approach is 
independent of the above mean field calculations for $T^*(\rho)$.
It can be used in connection with any method which yields a 
$T^* (\rho)$ curve.
The only requirement is that it would not cross the 
$T_c (\langle \rho \rangle)$ curve as some have proposed\cite{Tallon}. 
Thus the percolating approach could
be used with some experimental measurements or 
others theoretical methods to calculate 
$T^* (\langle \rho \rangle) $, like,
for instance, the calculations made with the Hubbard-Holstein
Hamiltonian\cite{Grilli}.
\section{Discussion}

There are several HTSC phenomena  which are not well understood
and  can be  explained with 
the percolating ideas, and with our model and
calculations:\\


1-ARPES measurements\cite{Renner,Harris} have revealed the 
anomalous behavior of the zero temperature gap $\Delta_0(\langle \rho \rangle)$
which decreases steadily with the
doping $\langle \rho \rangle$ although $T_c$ increases by a factor of 2 for
their underdoped samples.
This overall relation between $\Delta_0(\langle \rho \rangle)$ and $T_c$ is
totally unexpected since it is well known that normal superconductors
have a constant value for the ratio $2\Delta_0/k_BT_c$, being
3.75 for usual isotropic order parameter and 4.18 for
$d_{x^2-y^2}$ wave solution\cite{Maki}. 

\begin{figure}
\includegraphics[width=8cm]{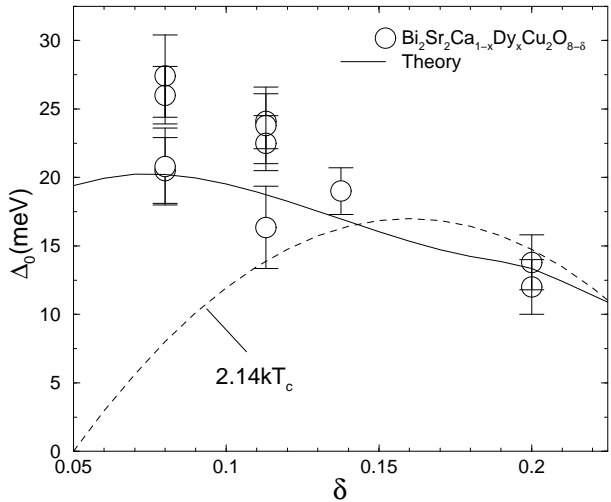}
\caption{The zero temperature gap for 9 samples as measured by Harris et
al\cite{Harris} and  our calculations.}
\end{figure} 
At low temperatures, since the superconducting phase
percolates through different regions, each one has a 
given $\Delta_0(r)$. Thus, tunneling and ARPES
experiments detect the largest gap present in the compound.
Consequently, $\Delta_0(\langle \rho \rangle)$ must be correlated with the onset
of vanishing gap $T^*(\langle \rho \rangle)$ which is the 
largest superconducting
temperature in the sample, and should not be
correlated with $T_c(\langle \rho \rangle)$. As we show in
Fig.5, correlating the values plotted in Fig.3b for $T^*(\langle \rho \rangle)$
with $\Delta_0(\langle \rho \rangle$), we are able to give a reasonable
fit for the data\cite{Renner,Harris} on $Dy-BSCCO$
and explains the different energy scales pointed out by
several others authors.

2-Transport experiments have also been used to study the pseudogap\cite{TS}.
The underdoped and optimum doped high-$T_c$ oxides have a linear
behavior for the resistivity  in the normal phase
up to very high temperatures.
However, at $T^*$ there is a deviation from the linear behavior and
the resistivity falls faster with decreasing temperature\cite{TS,Oda}.
This behavior can be understood by 
the increasing of superconducting cluster numbers and size, as the
temperatures decreases below $T^*$. Each superconducting cluster produces
a short circuit which decreases the resistivity below the
linear behavior between $T^*$ and  $T_c$. 
To obtain a quantitative fitting of this effect, we are presently working
on a simulation for the resistivity of a linear metallic medium
with short circuited regions which varies with the temperature.
The linear behavior of the resistivity above $T^*$ may be
also be explained by a percolation procedure. The conductivity of 
inhomogeneous systems has been considered from
different percolative approaches in order to obtain the
correct temperature dependence\cite{Pike,AHL}. The temperature
functional form depends on how the percolation is achieved
in the system.

3-Several measurements made in the presence of a magnetic field seem to 
agree with the percolating scenario: The magnetotransport 
measurements\cite{lake} 
in a $La_{2-x}Sr_xCuO_4$ film just above the irreversibility line is
in agreement with the notion that the magnetic field penetrates
partially in the superconducting regions and destroys some
of the superconducting clusters in the film. We have already mentioned
the recent measurements of magnetic domains  above $T_c$
which has been interpreted as a diamagnetic precursor to the
Meissner state, produced by performed pairs in underdoped
$La_{2-x}Sr_xCuO_4$ thin films\cite{IYS}. The existence of
superconducting clusters between $T^*$ and  $T_c$ easily
explains the appearance of local diamagnetic or Meissner domains and,
if there is a temperature gradient in the sample, the
local flux flows and produces the dynamic flux flow state\cite{Xu}.

4-Another important consequence which follows is that
the superconducting pairing mechanism should be more easily investigated by 
experiments performed mainly at $T^*$.
A such experiment was accomplished by Rubio
Temprano et al\cite{Temprano}, which measured a large isotope
effect  associated with $T^*$  and an almost
negligible isotopic effect associated with $T_c$
in the slightly underdoped $HoBa_2Cu_4O_8$ compound.
The results strongly support the fact that
electron-phonon induced effects are present in the superconducting
mechanism associated with $T^*$.
Bussmann-Holder et al have also calculated
$T^*$ as function of a phonon induced gap\cite{Annette}.

In order to gain further insight on
the nature of the pair potential, we have measured  the resistivity under
hydrostatic pressure on an
optimally doped $Hg_{0.82}Re_{0.18}Ba_2Ca_2Cu_3O_{8+\delta}$\cite{prbrap}
sample.  The  data
indicate a linear increase of $T^*$ with the pressure at the same
rate as $T_c$. In the
context of our theory, this result might be in agreement 
with the phonon induced mechanism: the inhomogeneities 
local charge densities in a given compound yield varying values
for the Fermi level, broadening $N(E_F)$\cite{OWK}. The applied pressure
on a cuprate with an inhomogeneous charge distribution
is also expected to broaden the density of states\cite{Angilella1}
$N(E_F)$, and the main effect of an applied pressure on  $T_c$
is an  increase of the phonon or Debye frequency.
This is seen through the linear
increase of $T^*$\cite{prbrap} which also provides
a very interesting physical explanation on the origin 
of the linear pressure induced
{\it intrinsic effect}\cite{Angilella1,Orlando,Jorge}, 
usually postulated to explain
the raise of $T_c$ above its maximum value.

5- The anomalous behavior found mostly in underdoped compounds is 
related with the temperature behavior of the Hall coefficient\cite{Ong}
$R_H$ and the Hall density $n_H$ which is proportional to
the average hole density $\langle\rho\rangle$. While in normal metals $n_H$ is
independent of the temperature, in the normal phase of many HTSC,
$n_H$ increases monotonically as the temperature increases and
saturates at high temperatures\cite{Ong,Ong2}.
This anomalous behavior can be explained by the existence
of superconducting islands above $T_c$ which gradually 
turn into metallic phase as the temperature is raised. Such superconducting
regions can be regarded as empty spaces in a disordered
hole-rich (metallic) and hole-poor (insulator)
background. Therefore, as the metallic
region increases, it  increases also the number of carrier. 
However, we have to point out  that the saturation seems
to occur near the room temperature, which is  much higher than 
some of the measured values for  $T^*$.

6- The pseudogap has also been detected through the specific heat coefficient
which undergoes changes below $T^*$. It is well known that the electronic 
specific heat $\gamma$ term of the normal phase is
a material dependent and temperature independent  constant.
However, several measurements have 
detected\cite{TS,Tallon,Loram}
the suppression in the $\gamma$ term in mostly underdoped
compounds of different families.  In the percolating approach, 
with the existence of superconducting
clusters below $T^*$, we can qualitatively explain such depression.
While the metallic region specific heat behaves as $\gamma T$, the
superconducting region contributes with terms proportional to
$\exp(-\Delta(r)/T)$, which decreases  much faster with $T$. As
the temperature falls below $T^*$ and the superconducting region
increases, its contribution to the specific heat becomes more important and
produces the overall downturn in the $\gamma$ coefficient. 
For overdoped samples, since $T^*$ is very close $T_c$, this effect
may be difficult to be detected and this is the suppression of the 
$\gamma$ term was not seem for overdoped compounds\cite{Tallon,Loram}.

7- Magnetization measurements above $T_c$ in oriented powder of 
$Y_{1-x}Ca_xBa_2Cu_3O_y$ has revealed a pattern\cite{Rigamonti} which
cannot be explained by the usual fluctuation of the magnetization
order parameter\cite{Tinkham}. Such pattern were interpreted by
the formation of superconducting regions above the percolating
threshold\cite{Rigamonti} as discussed in detail by the percolating
model above. We are presenting using the Ginsburg-Landau 
formalism\cite{Tinkham} in connecting of the hole charge distribution
introduced in Section II in order to provide a quantitavive
calculation for such anomalous induced diamagnetism. The results
will be published elsewhere together with a new analysis on the
behavior of $H_{c2}$ for high-$T_c$ superconductors.
\section{Conclusions}
We have demonstrated that  the percolating approach for
local pre-formed pairs due to an
inhomogeneous charge distribution  on the $CuO_2$ planes
provides a  novel and general  approach 
to the phase diagram of the 
high-$T_c$ cuprates superconductors. The pseudogap is regarded as 
the largest superconducting gap among the  superconducting regions in an
inhomogeneous compound. The critical temperature $T_c$ is
the maximum temperature at which these superconducting regions
percolate. We have shown, through the calculations presented
in this paper, that this approach is suitable to reproduce 
the measured $T_c$ and $T^*$ phase diagrams  
for several cuprates using a phenomenological real charge distribution
drawn from the STM/S data. 

One of the advantages of our approach is that the pair 
formation and the phase coherent superconducting state are
studied independently. The  phase coherence is attained by
percolation. On the other hand,  the pair formation 
was studied by a mean field BCS like method
in order to estimate the onset of superconducting gap 
$T^*$ for any doping level. This was the simplest method
and the calculations  could have been done
with other theoretical approaches. However, notice that the
pair formation mechanism is
independent of how the coherent percolative phase is reached. 
This  general approach is
in agreement with new trends and ideas\cite{Wang,Burgy}, which are
based on new data and new experimental facts that mostly
HTSC are inhomogeneous materials.

The method developed  provides also new insights  and 
introduces a new interpretation 
on several typical anomalous properties of HTSC: like the 
dependence of the zero temperature gap $\Delta_0(T)$ 
with the hole concentration, the downturn of the linear dependence
to the resistivity with the temperature, the linear dependence of
the hole concentration with the temperature, the suppression
of the specific heat coefficient $\gamma$, the dependence of $T^*$
with the pressure, and others new implications which are 
presently being studied and which will be
discussed in the future.

Financial support of CNPq and FAPERJ is gratefully acknowledged.
JLG thanks CLAF for a CLAF/CNPq pos-doctoral fellowship.

\end{multicols}
\end{document}